\documentclass[%
 aip,
 apl,%
 amsmath,amssymb,
 reprint,%
]{revtex4-1}

\usepackage{graphicx}
\usepackage{dcolumn}
\usepackage{bm}

\begin{document}

\preprint{AIP/123-QED}

\title[H. Shioya et al.]{Gate tunable non-linear currents in bilayer graphene diodes
}

\author{Hiroki Shioya}
 \email{shioya@meso.t.u-tokyo.ac.jp}
 \affiliation{Department of Applied physics, University of Tokyo, Japan.}
\author{Michihisa Yamamoto}
 \affiliation{Department of Applied physics, University of Tokyo, Japan.}
\author{Saverio Russo}
 \affiliation{Centre for Graphene Science, CEMPS, University of Exeter, United Kingdom.}
\author{Monica F. Craciun}
 \affiliation{Centre for Graphene Science, CEMPS, University of Exeter, United Kingdom.}
\author{Seigo Tarucha}
\affiliation{Department of Applied physics, University of Tokyo, Japan.}

\date{\today}

\begin{abstract}
Electric transport of double gated bilayer graphene devices is studied as a function of charge density and bandgap. A top gate electrode can be used to control locally the Fermi level to create a pn junction between the double-gated and single-gated region. These bilayer graphene pn diodes are characterized by non-linear currents and directional current rectification, and we show the rectified direction of the source-drain voltage can be controlled by using gate voltages. A systematic study of the pn junction characteristics allows to extract a gate-dependent bandgap value which ranges from 0 meV to 130 meV.
%
\end{abstract}

\pacs{Valid PACS appear here}
\maketitle

%



Graphene -a single layer of carbon atoms- is the thinnest known conductor \cite{Novoselov2005, Zhang2005} with a room temperature charge carrier mobility considerably higher than silicon (i.e. more than $100,000 cm^2/Vs$). One of the major challenges in graphene-based electronics is that charge carrier conduction cannot be simply switched off by means of gate voltages because of the lack of a bandgap \cite{no-gap}. While chemical functionalization constitutes a valuable approach to engineer a bandgap \cite{Chemical}, bilayer graphene offers an alternative solution to this problem with a gate tunable bandgap \cite{McCann2006, Castro2007}. Indeed, bilayers can be continuously driven from the semimetal to insulator state simply by means of gate voltages. This remarkable property paves the way towards bilayer-based transistor and diode applications with large on/off ratio of the current.

The observation of an electric field tunable bandgap in bilayer graphene was reported in infrared optical spectroscopy \cite{Zhang2009,Mak2009} and a method, which is different from that \cite{Xia2010}. However, a direct measure of this energy-gap in electrical transport experiments has so far been elusive due to the presence of disorder induced sub-gap states dominating the low energy transport properties. Indeed, systematic electrical transport experiments demonstrated that the temperature dependence of the bilayer graphene resistance for sub-gap energies is successfully described by mainly two parallel electrical transport channels \cite{Oostinga2008, Miyazaki2010, Taych2010, Russo2009, Monica2011}. They are caused by variable range hopping (and by nearest neighbor hopping especially in the lower temperature region than variable range hopping \cite{Taych2010}) and by thermally activated transport over the bandgap, respectively. The formation of bilayer graphene pn junctions has attracted considerable theoretical interest \cite{Joglekar,Cheng2010,Saisa-ard} and it is a key element in graphene-based electronics. Though up to date there is only one experimental report on pn-junction devices, no evidence for a rectified current has yet been reported leaving graphene-based electronics still a pure academic exercise \cite{PreviousPNex}.

In this letter we show that the current-voltage (I-V) characteristics of the pn junction devices exhibit a rectifying behavior with the threshold given by the bandgap rather than the disorder induced sub-gap states.Then we analyze the I-V characteristics to derive the electric field tunable bandgap of bilayer graphene. We finally extract a gate-dependent bandgap value ranging from 0 meV to 130 meV in our double gated bilayer graphene devices.

The graphene devices are fabricated by mechanical exfoliation of Kish graphite on p-doped Si wafer coated by 285nm $SiO_{2}$, which acts as a uniform back gate. We have selected bilayer graphene flakes by analyzing the intensity of the green light of optical micrograph pictures of different flakes \cite{Monica2009}. Subsequently, we fabricated electrodes and local top gates by electron beam (EB) lithography, evaporation and lift-off process of respectively Ti/Au (25 nm/65 nm) for the electrodes and SiO2/Ti/Au (150 nm/25 nm/32 nm) for the top gates. Figure 1(a) shows an optical micrograph of a typical double gated pn-junction device.

\begin{figure}
\includegraphics[width=0.31\textwidth]{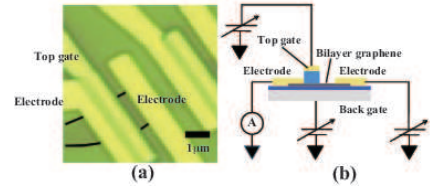}
\caption{\label{fig:epsart} (Color online) (a) Optical micrograph of a measured device: Black lines show the edge of bilayer graphene flake.
(b) Schematic diagram of our measurement system.
}
\end{figure}

\begin{figure*}
\includegraphics[width=0.73\textwidth]{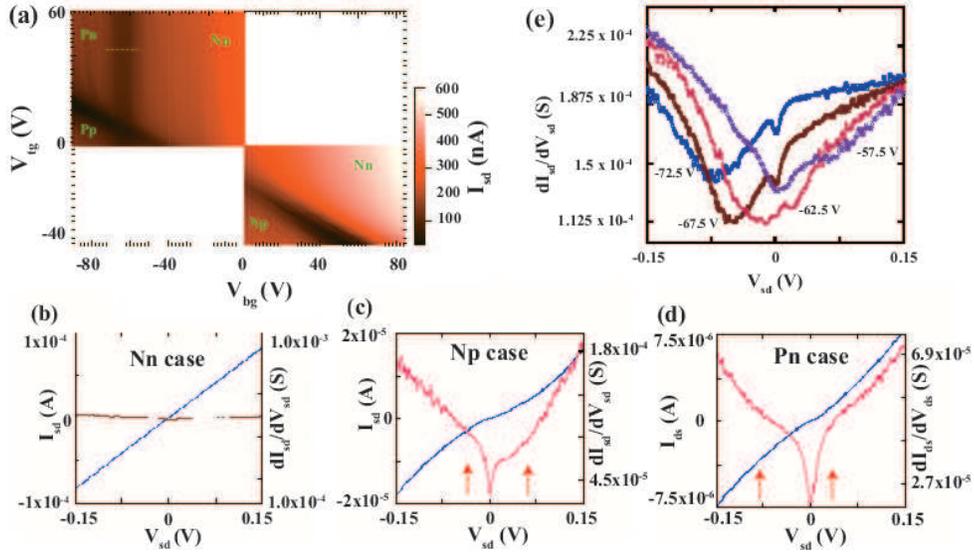}
\caption{\label{fig:wide} (Color online) (a) Color coded plot of $I_{sd}$ \textit{versus} $V_{tg}$ and $V_{bg}$ at T=4.2K. I-V characteristic of the Nn junction with $V_{tg}=-35V and V_{bg}=80V$ (b), Np junction with $V_{tg}=-45V and V_{bg}=55V$ (c), Pn junction with $V_{tg}=20V and V_{bg}=-93.45V$ (d), respectively. (e) Differential conductance vs. $V_{sd}$ plotted for a range of back gate voltages with $V_{tg}=40V$. The back gate voltages correspond to measurements at points on the dotted line in (a). Minimum points are shifted indicating the shift of the quasi-Fermi level. The number attached to each curve shows the applied $V_{bg}$ value.
 }
\end{figure*}

The double-gated geometry allows us to independently control the Fermi level and the perpendicular electric field through the bilayer graphene. The polarity of the charge carriers -i.e. electrons (n) or holes (p)- can be controlled by tuning the Fermi level position via both top gate and back gate voltages. This property is used to create controllable pn-junctions. On the other hand, the external perpendicular electric field breaks the energetical symmetry between the two planes of the bilayer graphene resulting in the opening of a bandgap \cite{McCann2006}.

Here we use more than 5 different double gated pn-junction devices to the study of the electrical properties in a voltage-bias configuration (see Figure 1(b)) with top and back gate voltages as parameters and at a temperature of 4.2K. In all these bilayer pn-junction devices we observed similar I-V characteristics. Here we discuss the representative data obtained from one of them. To estimate the mobility (${\mu}$) of the charge carriers, we use the value of the conductance per square measured at high gate voltage i.e., $G_{\square}=ne {\mu}$, from which we have subtracted the contact resistance between bilayer graphene and Ti/Au \cite{Russo2010}. The density of charge carriers is given by the known capacitance to the gate, and using the known planar capacitance in the device channel we obtain $\mu = 650 cm^2 /V s$ at 4.2K. Thus the typical electron mean free path in our devices is about 16.5 nm, which is much shorter than the device channel. Therefore, the electron transport in our devices is diffusive.



Figure 2(a) shows a color coded plot of source-drain current ($I_{sd}$) measured for a constant source-drain voltage $V_{sd}=1mV$ as a function of top- and back-gate voltages, $V_{tg}$ and $V_{bg}$, respectively. The $I_{sd}$ shows two minima, one corresponding to the charge neutrality point of the double gated region -which depends on both $V_{tg}$ and $V_{bg}$- and the other corresponding to that of the region not covered by the top-gate -i.e. independent of $V_{tg}$. These two minima cross at $V_{tg} = 8V$ and $V_{bg} = -64V$ which is the global neutrality point for the whole device. Therefore, by adjusting the gate voltages we can prepare four different polarity regions, i.e. Pp, Pn, Np and Nn as indicated in Figure 2(a). For instance, for $V_{bg} {\le} -64V$ in the region of the flake without top-gate the Fermi level is in the valence band (P-doping). By application of the specific top-gated voltages the Fermi level of the double gated region can be brought from the valence (p-doping ) to the conduction band (n-doping) realizing a transition of a Pp to Pn-junction.

The I-V characteristics of double gated bilayer graphene devices measured for the Nn, Np and Pn conditions are shown in Fig. 2 (b), (c) and (d), respectively. For the Pp (or Nn) junction we  observe linear I-V characteristics, whereas for the Pn (or Np) junctions, which are realized by the gate voltages near charge neutral points we consistently observe nonlinear I-V characteristics. This nonlinearity becomes more pronounced when the Fermi level is set in the opened bandgap of the p-region. Correspondingly, the non-linear I-V characteristic is also reflected in the $dI_{sd}/dV_{sd}$ \textit{vs.} $V_{sd}$ characteristics (see Fig 2. (c) and (d)). The $dI_{sd}/dV_{sd}$ starts to increase progressively when $V_{sd}$ is tuned outside of the region, indicated with two arrows in Fig 2. (c) and (d): The $dI_{sd}/V_{sd}$ value is apparently reduced in the positive $V_{sd} (0mV {\le} V_{sd} {\le} 70mV)$ in Fig. 2 (c) and in the negative $V_{sd} (0mV {\ge} V_{sd} {\ge} -80mV)$ in Fig.2 (d), respectively. That is the rectified direction in the $V_{sd}$ axis, which changes with the polarities in the channel of the device, e.g. negative $V_{sd}$ for Pn and positive $V_{sd}$ for Np. This observation implies that in our devices the rectification direction of $V_{sd}$ can be controlled by selecting the appropriate gate voltage configuration. In addition, in a higher electric field -i.e. wider bandgap- we observed an even more pronounced non-linearity in the I-V characteristics(see Fig. 3 (c)). The zero-bias dip in conductance, see Fig. 2c and d, is likely to originate from both contact resistance and disorder induced sub-gap states. The energy range associated with this zero-bias dip is much smaller than the estimated energy gap, see Figs. 2(c)-2(e), highlighting a non-leading role of the contact resistance in these devices.







Contrary to conventional semiconductor pn-junctions, there is no so-called built in potential at a bilayer graphene pn interface. In the bilayer graphene devices the fringing field from the side edge of the top gate smears out the pn interface. Figure 3 (a) shows a simplified schematic of the band-bending at a pn bilayer interface. Here the bandgap is assumed to act as a potential barrier for carriers and to generate non-linear currents. We confirmed the validity of this model from measurements of the $V_{bg}$ dependence of the differential conductance, see figure 2 (e). In this figure the Fermi level is tuned from below to above the bandgap by changing $V_{bg}$ from -72.5V to -57.5V. The position and the width in $V_{sd}$ of the differential conductance depend on $V_{bg}$. This observation suggests that the non-linear current is a consequence of transport through a Pn interface where the gap acts as a potential barrier for charge carriers.

\begin{figure}
\includegraphics[width=0.48\textwidth]{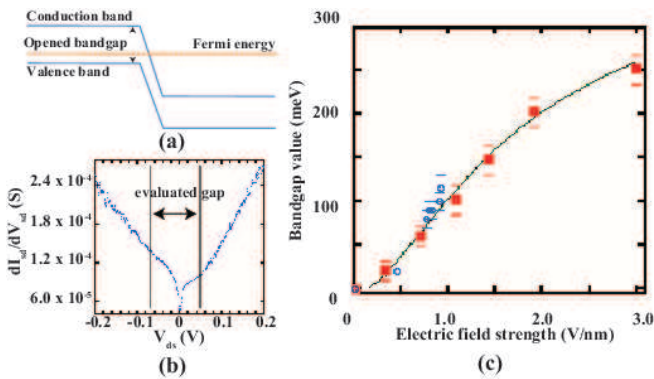}
\caption{\label{fig:epsart} (Color online) (a) Schematic of band structure at an interface when the Fermi level is set in the opened bandgap: The differential conductance would show different slopes outside of the gap in the differential conductance curves when carriers are driven by the $V_{sd}$ voltage. Carrier conduction would be suppressed when carriers experience the opened bandgap.
(b) Analysis of a differential conductance curve ($V_{tg}=-40V, V_{bg}=37.5V$): The point at which the slope of derivative conductance starts to change corresponds to the edge of the opened bandgap. The product of difference between 2 inflection points of $V_{sd}$ and e of elementary charge corresponds to the opened bandgap value.
(c) Evaluated bandgap values vs. the applied electric fields. Circles are the evaluated values of our analysis. Squares are the values of previous optical measurements(see Ref. 7) and continuous curve is the theoretical calculation in the previous work(see Ref. 7).
}
\end{figure}

Within the proposed band-bending model we estimate the value of the electric field induced bandgap opened in bilayer graphene. A typical $dI_{sd}/dV_{sd}$ \textit{vs.} $V_{sd}$ is shown in figure 3 (b). We start by analyzing the differential conductance curves with the quasi-Fermi level set as depicted in figure 3 (a), i.e. in the bandgap under the top gate (left region) and in the conduction band outside the top gate (right region). When the quasi-Fermi level  reaches the bottom (top) of the conduction (valence) band under application of $V_{sd}$ an inflection point appears in the $dI_{sd}/dV_{sd}$ \textit{vs.} $V_{sd}$ curves (see figure 3 (b) caption). For the forward $V_{sd}$, the carrier conduction will increase when the incident energy of carriers overcome the energy of the potential barrier. For the backward $V_{sd}$, the carrier conduction will increase when the carriers in the valence band in the P-region start to flow. Therefore, the two inflection points correspond to the situations where the quasi-Fermi level is located at the upper edge and the lower edge of the gap, respectively. The $V_{sd}$ range between the two inflection points in figure 3 (b) is 110mV, we can therefore infer that the opened bandgap value is 110meV for $V_{tg}=-40V$ and $V_{bg}=37.5V$. We apply this analysis for the data with combinations of gate voltages and derive the bandgap value ranging from 0meV to 130meV, see blue circles in figure 3 (c). Furthermore, we compare our estimates to those previously obtained from infrared optical spectroscopy \cite{Zhang2009}, see red squares in figure 3 (c). The magnitude and the overall electric field dependence of the bandgap obtained from our electrical transport experiments are both in good agreement with those of the previous work \cite{Zhang2009}.

In conclusion, we demonstrated that local electrostatic doping of bilayer graphene can be used to form pn diodes with nonlinear rectified current-voltage characteristics. The rectification behavior can be continuously tuned by means of gate voltages. From the analysis of the pn junction current-voltage characteristics we estimate a gate-dependent bandgap as large as 130mV, which is in agreement with previous optical studies. Our experimental findings demonstrate a step forward to future bilayer graphene diode applications.

(Acknowledgement) H.S. acknowledges financial support from GCOE for Phys. Sci. Frontier and from Project for Developing Innovation Systems, MEXT, Japan. 
M.Y. acknowledges financial support from Grant-in-Aid for Young Scientists A (no. 20684011). 
S.T. acknowledges financial support from JST Strategic International Cooperative Program (GFG-JST and EPSRC-JST).
S.R. and M.F.C. acknowledge financial support from EPSRC (Grant No. EP/G036101/1 and no. EP/J000396/1) and from the Royal Society Research Grant 2010/R2 (Grant no. SH-05052) and 2011/R1 (Grant No. SH-05323).

\end{document}